\documentclass[preprint2]{aastex631}

\usepackage{CJK}
\shorttitle{The structure of open clusters}
\shortauthors{Zhong et al.}
\graphicspath{{./}{figures/}}
\begin{document}
\begin{CJK*}{UTF8}{gbsn}
\title{New insights into the structure of open clusters in the Gaia era}

\author[0000-0001-5245-0335]{Jing, Zhong (钟靖)}
\affiliation{Key Laboratory for Research in Galaxies and Cosmology, Shanghai Astronomical Observatory, Chinese Academy of Sciences,80 Nandan Road, Shanghai 200030, China. jzhong@shao.ac.cn}

\author{Li, Chen (陈力)}
\affiliation{Key Laboratory for Research in Galaxies and Cosmology, Shanghai Astronomical Observatory, Chinese Academy of Sciences,80 Nandan Road, Shanghai 200030, China}
\affiliation{School of Astronomy and Space Science, University of Chinese Academy of Sciences, No. 19A, Yuquan Road, Beijing 100049, China}

\author{Yueyue, Jiang (蒋悦悦)}
\affiliation{Key Laboratory for Research in Galaxies and Cosmology, Shanghai Astronomical Observatory, Chinese Academy of Sciences,80 Nandan Road, Shanghai 200030, China}
\affiliation{School of Astronomy and Space Science, University of Chinese Academy of Sciences, No. 19A, Yuquan Road, Beijing 100049, China}

\author{Songmei, Qin (秦松梅)}
\affiliation{Key Laboratory for Research in Galaxies and Cosmology, Shanghai Astronomical Observatory, Chinese Academy of Sciences,80 Nandan Road, Shanghai 200030, China}
\affiliation{School of Astronomy and Space Science, University of Chinese Academy of Sciences, No. 19A, Yuquan Road, Beijing 100049, China}

\author{Jinliang, Hou (侯金良)}
\affiliation{Key Laboratory for Research in Galaxies and Cosmology, Shanghai Astronomical Observatory, Chinese Academy of Sciences,80 Nandan Road, Shanghai 200030, China}
\affiliation{School of Astronomy and Space Science, University of Chinese Academy of Sciences, No. 19A, Yuquan Road, Beijing 100049, China}

%% Mark off the abstract in the ``abstract'' environment. 
\begin{abstract}
With the help of Gaia data, it is noted that in addition to the core components, there are low-density outer halo components in the extended region of open clusters. In order to study the extended structure beyond the core radius of the cluster ($\sim$ 10 pc), based on Gaia EDR3 data, taking up to 50 pc as the searching radius, we use the pyUPMASK algorithm to re-determine the member stars of the open cluster within 1-2 kpc. We obtain the member stars of 256 open clusters, especially those located in the outer halo region of open clusters. Furthermore, we find that the radial density profile in the outer region for most open clusters deviates from the King’s profile. In order to better describe the internal and external structural characteristics of open clusters, we propose a double components model for description: core components with King model distribution and outer halo components with logarithmic Gaussian distribution, and then suggest using four radii ( $r_c$, $r_t$, $r_o$, $r_e$) for describing the structure and distribution profile of star clusters, where $r_t$ and $r_e$ represent the boundaries of core components and outer halo components respectively. Finally, we provide a catalog of 256 clusters with structural parameters. In addition, our study shows the sizes of these radii are statistically linear related, which indicates that the inner and outer regions of the cluster are interrelated and follow similar evolutionary processes. Further, we show that the structure of two components can be used to better trace the cluster evolution properties in different stages.

\end{abstract}

%% Keywords should appear after the \end{abstract} command. 
%% The AAS Journals now uses Unified Astronomy Thesaurus concepts:
%% https://astrothesaurus.org
%% You will be asked to selected these concepts during the submission process
%% but this old "keyword" functionality is maintained in case authors want
%% to include these concepts in their preprints.

%% From the front matter, we move on to the body of the paper.
%% Sections are demarcated by \section and \subsection, respectively.
%% Observe the use of the LaTeX \label
%% command after the \subsection to give a symbolic KEY to the
%% subsection for cross-referencing in a \ref command.
%% You can use LaTeX's \ref and \label commands to keep track of
%% cross-references to sections, equations, tables, and figures.
%% That way, if you change the order of any elements, LaTeX will
%% automatically renumber them.
%%
%% We recommend that authors also use the natbib \citep
%% and \citet commands to identify citations.  The citations are
%% tied to the reference list via symbolic KEYs. The KEY corresponds
%% to the KEY in the \bibitem in the reference list below. 

\section{Introduction}
\label{sec:intro}
The structural composition and the boundary of open clusters have great significance in the study of open clusters, which directly affect the measurement results of other parameters such as mass function, binary fraction, mass ratio distribution, evolution time scale, and dynamic mass of open clusters. It also affects our understanding of the formation, evolution, and dissolution of star clusters.

As early as the early 20th century, \citet{1916CMWCI.117....1S} and \citet{1918PA.....26Q...9T} speculated that the generally observed open star cluster may only be its high-density core component, and there may be external low-density halo/corona components in its outer region. Although the density of member stars in the outer region is very low, since the scale is 5-10 times larger than that of the core region, the total number of member stars may still be several times or even ten times larger than those in the core region \citep{1969SvA....12..625K}.

In general, the traditional boundary (core area) of open clusters is a location where the density of member stars is approximately equal to that of background field stars \citep[r2,][hereafter K13]{2013A&A...558A..53K}, while field stars are dominant beyond this region. In observation, because the number density of member stars in the outer region is significantly lower than that of the background field star, it is difficult to effectively distinguish the member stars from the field stars without additional information.  This also means that high-precision kinematic data is of key importance to identify member stars, especially in the outer region. 

Fortunately, Gaia data provide us with high-precision astrometric data, which also means one can obtain highly reliable results of membership identification for further studying cluster properties, such as the extended structure of clusters. With the help of Gaia data, many tidal structures of old open clusters have been discovered: \citet{2019A&A...621L...2R} and \citet{2019A&A...621L...3M} confirmed the existence of a huge tidal structure of more than 170 pc in Hyades for the first time; based on similar methods, \citet{2019A&A...627A...4R} further studied Praesepe and found that the cluster has a tidal tail structure of 165 pc; \citet{2019A&A...627A.119C} also found that there is a tidal structure of more than 50 pc around the famous cluster M67; \citet{2020ApJ...889...99Z} found that the younger cluster Blanco\_1  also has an extensional structure of more than 50 pc; \citet{2019AJ....157..115Y} investigated the dynamical status of the nearby cluster Rupecht\_147 and found the existence of prominent tidal tails as well as an extended corona; in addition, \citet{2022RAA....22e5022B} reports an elongated tail structure of the newly discovered open cluster COIN-Gaia 13, with a whole length of about 270 pc.

Using Gaia data, \citet{2019A&A...624A..34Z} investigate member stars of young double clusters NGC\_869 and NGC\_884 within a radius of 7.5 degrees. It is found that there is an obvious outer halo structure, which is mainly distributed in the radius of 50 pc. This discovery confirms for the first time that very young clusters also have an extended structure. The newly discovered outer halo region is 6-8 times larger than the traditional boundary ($r_2$) defined by K13. This study also shows that, with the help of Gaia DR2, the extended outer halo structure of clusters within a distance of about 2kpc can be well distinguished from the background field stars. 

Although outer halo (or corona) structures and tidal tails have been confirmed in the extended region of the open clusters, the two structures have different characteristics and origins from the perspective of cluster dynamic evolution. According to the numerical experimental results of \citet{2008ARep...52..467D}, after 150 Myrs of the open cluster evolution, the total extent of the tidal tails can reach 1.2 kpc. \citet{2020ARep...64..827T} develop a concept of the stellar streams forming due to the decay of star clusters, which can finally form the tor-like structures in the Galactic disc. It can be seen that the tidal tail of star clusters is mainly formed by the motion and dynamical evolution  in the Milky way. On the other hand, according to the dynamical study by \citet{2014ARep...58..906D}, the corona of the cluster is a long-lived structure outside the zero-velocity surface, whose temporal stability is determined by the existence of the periodical retrograde orbits  (with periods comparable to the mean lifetime of the cluster) and the trajectories close to such orbits. Therefore, the outer halo component can be regarded as an important part of star clusters.

The study of the outer structure of clusters is crucial in understanding the formation and evolution of star clusters. The existence of tidal structure in the extended region revealed by Gaia data can help one better understand the dynamic evolution of star clusters and further reveal the evolution time scale of star clusters in the Milky way \citep{2019ARA&A..57..227K}. It is worth noting that young star clusters also have originally formed outer halo structures \citep{2019A&A...624A..34Z,2021A&A...645A..84M}. This discovery provides us a new insight to study the formation mode as well as the initial mass function (IMF) of star clusters in the giant molecular cloud: from the core region to the outer halo region, the different interstellar environments can be used to study the influence of different molecular cloud densities on the IMF, the key factors leading to the power-law form of the IMF, and the main physical processes affecting the IMF \citep{2010A&A...518L.106K,2015A&A...584A..91K,2016MNRAS.458.1671K,2019A&A...629L...4A}. In particular, compared with the core region, these primordial outer halo structures retain the initial formation state of the cluster to the greatest extent and will be better used to study the low-mass end of the IMF.

In order to explore the question of whether open clusters generally have an outer halo structure, and further study  outer halo structures more comprehensively and systematically, we chose a larger area for member stars searching and tried to establish a cluster sample with extended outer halo structures. Our primary goal is to identify outer halo structures and try to re-describe the radial density profile of clusters considering the extended structure. The paper is organized as follows: in Section~\ref{sec:dat}, we mainly introduce the discovery of extended structures in a large sample of clusters and how to establish a double components model for describing the radial density profile of clusters; in Section~\ref{sec:res}, we develop characteristic radii to describe the spatial structure of clusters and further discuss the characteristics of these radii. Furthermore, we investigate the relation of cluster size along with age, while it can be used to infer the evolution of open clusters. In Section~\ref{sec:sum}, we summarise our results.  

\section{Data analysis}
\label{sec:dat}
\subsection{Sample selection}
In order to fully reveal the extended structure of open clusters, we searched for member stars in the outer region of clusters. As the early stage of Gaia third data release \citep{2021A&A...649A...1G}, the Gaia EDR3 catalog updates the astrometry and photometry for 1.8 billion sources brighter than 21 mag in G-band, with higher precision than Gaia DR2. The cluster coordinates and fundamental parameters (mean proper motions, distance, age and extinction) are mainly referred to the latest cluster catalog provided by \citet[here after CG20]{2020A&A...640A...1C}. In the work of CG20, the radius of the searching cone field for each cluster mainly refers to the literature results of \citet{2002A&A...389..871D} and K13, the procedure of membership determination mostly only focus on the core members. It is necessary for us to expand the search radius to identify the members of more extended areas, and further define the size of the cluster containing new extended components. 

We first select clusters with $logt> 7$. In this work of investigating the extended components of clusters, we exclude very young clusters because they are more likely to be suffered from serious optical extinction, which means that the less completeness of their member stars compared with the ordinary cluster, especially in the outer region of the cluster. Then, to ensure the reliability of identifying member stars in the outer region, we limit the cluster distance in the range from 1000 pc to 2000 pc. The astrometric precision of Gaia EDR3 in this distance range is still high enough to well distinguish between cluster members and field stars. Finally, we select 434 star clusters from CG20 for further membership identification, especially in the extended structure of their outer region. 

For each star cluster, we adopt cluster celestial coordinates as search center and use the Python Astroquery package \citep{2018AJ....156..123A} to retrieve sources within the cluster searching radius through the Gaia archive\footnote{https://gea.esac.esa.int/archive/}. Considering the typical size for most giant molecular clouds of about 200 pc, we adopt the searching radius of cluster members with 100 pc, to cover the whole molecular cloud region in which cluster members may have formed. For clusters with a distance between 1000 pc and 2000 pc, the corresponding angular searching radius is from 5.7$^{\circ}$ to 2.8$^{\circ}$. Then, we both apply the quality filters proposed by \citet{2018A&A...616A..17A} and the magnitude cut of G-band brighter than 18 mag to exclude sources without high quality in the 5-parameter astrometric solution. In order to reduce the contamination of field stars in the outer region, for each cluster, we further exclude sources whose proper motion locate outside the criteria circle which is centered on their expected average proper motion ($\mu_{\alpha^{\ast}}$, $\mu_{\delta}$) with a radius of 3 times the total proper motion dispersion ($\sigma_{\mu}$). The expected average proper motion ($\mu_{\alpha^{\ast}}$, $\mu_{\delta}$), as well as the corresponding dispersion ( $\sigma_{\mu_{\alpha^{\ast}}}$,$\sigma_{\mu_{\delta}}$) were provided by CG20, while the total proper motion dispersion can be estimated as $\sigma_{\mu}=\sqrt{\sigma^{2}_{\mu_{\alpha^{\ast}}}+\sigma^2_{\mu_{\delta}}}$.

Subsequently, we used an improved unsupervised clustering package called pyUPMASK \citep{2021ascl.soft01016P} to perform the membership identification of each star cluster. It is a package based on the UPMASK approach \citep{2014A&A...561A..57K} but compiled with the Python language and included several key enhancements, such as providing multiple clustering methods. The UPMASK is a robust approach that has been successfully applied in many works of cluster membership identification \citep{2018A&A...618A..93C,2019A&A...624A.126C,2020A&A...640A...1C,2019A&A...627A.119C}.

In the process of using pyUPMASK to determine membership probabilities, we use the K-means  clustering algorithm to determine the clumps in three-dimensional astrometric parameters ($\mu_{\alpha}$, $\mu_{\delta}$, $\varpi$) and choose the default value of pyUPMASK package nclust = 25 as the reasonable average number of stars per clustering subsamples. According to the assigned membership probability results, we select stars whose membership probability is greater than 0.7 as member stars, then we pass through the visual inspection for each cluster to make sure the reliability of our membership identification. In this work, we define high-quality clusters as both presenting a clear main sequence pattern in the color-magnitude diagram, an obvious single dense core in the celestial coordinate diagram, and a relatively clumpy distribution in the proper motion and parallax diagram. After the visual inspection, we retain 256 clusters as high-quality clusters with reliable members to perform the structure distribution analysis. Among the excluded 178 clusters which were classified as unreliable results, the majority of them were unable to present enough characteristics of the open cluster in parameter space because of the misidentification of cluster members. The misidentification is mainly caused by the small number of cluster members and/or too many field stars in the extended searching region, which may cause the contamination rate to increase to more than 60\%  \citep{2014A&A...561A..57K}. In addition, since we mainly focus on the extended structure of single clusters, we also excluded clusters that have multiple density cores in the spatial space ($\sim$ 15\% in 178 unreliable clusters), and clusters which have inapparent core regions because of the too few member stars ($\sim$ 10\% in 178 unreliable clusters).   

\subsection{Radial density profile}
A spatial density profile is a common tool to reveal the structure of star clusters and is also used for cluster size determination\citep{2012MNRAS.423.1940C,2013AJ....145...37S}.  In general, considering most of dense star clusters present approximate symmetrical structures, their spatial density profile can be dimensionally reduced to the radial density profile (RDP). To describe the RDP of a star cluster, the \citet{1962AJ.....67..471K} model is widely used for approximation of this kind of stellar distribution. Although the King model is an empirical density law derived from the RDP of globular clusters, it is still a good approximation function and is widely used for describing the RDP of many open clusters. The profile is described as:
\begin{equation}
\label{king}
    f (r)=k\cdot \left( \frac{1}{\sqrt{1+(r/r_{c})^2}}-\frac{1}{\sqrt{1+(r_{t}/r_{c})^2}}\right)^2
\end{equation}
Where $k$ is a constant, $r_c$ and $r_t$ are the core and tidal radius respectively. If $r$=$r_t$, the star density of cluster members is equal to zero, which means there are no cluster members outside this region. If $r$ $\gg$ $r_c$, the star density of cluster members tends to be a constant, and is often considered to be approximately equal to the density of surrounding field stars \citep{2013A&A...558A..53K,2016MNRAS.456.3757S,2021RAA....21...45Q}. In general, the ratio of $r_t$/$r_c$ is of the order of 10 in most open cluster  and the mean core radius is $r_c$=1.8 pc \citep{2013A&A...558A..53K}. Obviously, in the traditional study of the clusters, since the RDP fitting is mainly focused on a limited spatial area ( usually within a few dozen pc) and with a higher density than surrounding field stars, the approximation of the king's profile could usually perform a satisfactory description. 

However, in the Gaia era, because of the great improvement of astrometric data quality, more and more extended structures with large spatial distribution in the cluster were discovered \citep[e.g.,][]{2019A&A...624A..34Z,2019A&A...627A...4R,2020ApJ...889...99Z,2022A&A...659A..59T}. We now know that, for some open clusters, their member stars do not only include the core component but also the outer halo ( or corona) component. Since the outer halo members present a much more extended distribution than the core members, understanding of the structure and boundary of a cluster should be significantly changed. Moreover, although the number density of halo components is lower than the core components, a large spatial distribution of them could even provide a 10 times larger number of members than the core stars\citep{1969SvA....12..625K}. Considering that the outer halo members are also an important part of the cluster, their contribution to the cluster structure parameters can no longer be ignored.

To investigate the modified RDP of a cluster including outer halo members, we adopt the two-dimensional Gaussian kernel density estimation (KDE) to obtain the probability density distribution function of member stars ( P $>$ 0.7 ) in the cluster spatial position. It is a standard KDE that uses an automatic bandwidth estimator with Scott rule \citep{2015mdet.book.....S} for bandwidth calculation. Since the spatial number density distribution of member stars is mainly unimodal distribution, the bandwidth determined by the Scott rule can be considered the optimal bandwidth value to avoid over smoothing or under smoothing. Then, in the cluster region, we use a 200*200 grid to calculate the probability density distribution of member stars at each spatial point and  choose the average position of the 40 highest density points as the center of the cluster. The grid of probability density distribution is normalized by the maximum probability density. The sampling interval of RDP is determined by the interval of normalized density contour: when the probability density is between 1 and 0.1, the interval of isodensity is 0.05; When the probability density is less than 0.1, the interval of isodensity  is 0.025. For the grid points in each ring of the density contour (for example, the probability density is between 0.1 and 0.15), calculate their average radius, average probability density and the corresponding standard deviations, and use these values to represent the sampling points and error bars of RDP in the X direction (radius from the center of the cluster) and Y direction (normalized probability density). 

It is note that for most of clusters, the modified RDP can be divided into two parts: the inner part with small radial dispersion, and the outer part with great radial dispersion. This is because the distribution of core members in the inner region is usually dense and spherical symmetry  ( similar to the globular cluster distribution), while the distribution of density contours in the outer region is often asymmetric and presents a much gentle gradient. After adding the outer halo members in the cluster RDP, we note that the King model does not reproduce the modified RDP very well: 1) the $r_t$ will be derived to a very large value ( thousands of degrees) for some clusters, which is no longer represent a physical concept as King model described; 2) the overall fitting result often present a large deviation both in the inner and the outer region.   

Further analysis shows that the observed RDP can be divided into two parts: the inner part with dense distribution which can be well fitted by the traditional King model, and the outer part with extended distribution which reveals a large deviation with the King profile. It is noted that, \citet{2012ARep...56..623D} also found that the single King profile tends to underestimate the number of stars and proposed a double components model for the RDP of the open clusters. In addition to the cluster core, the corona model of a uniform sphere was constructed by \citet{2012ARep...56..623D}, and the surface density distribution of clusters is approximated by a double components model \citep{2016MNRAS.456.3757S}. Keeping these studies in mind, to perform a better description of the modified RDP, we assume that the core region is composed of core members that still consistent with the King model, and the outer region is composed of two components dominating with outer halo members. The double components model $F(r)$ for approximation of the modified RDP is
\begin{equation}
    F (r) = f (r) + g (r)
\label{comb}
\end{equation}
where $f(r)$ is the King model in Equation~\ref{king}, and $g(r)$ is a logarithmic  Gaussian function that described the RDP of outer halo members. In the outer region (r $>$ $r_c$), the RDP of outer halo members is
\begin{equation}
       g (r)  = \rho \cdot e^{- \frac{( \mathrm{ln} (r)-\mu)^2}{2\sigma ^2} }
\label{lognora}
\end{equation}
where $\rho$, $\mu$, and $\sigma$ are the free parameters for fitting. 

According to \citet{2014ARep...58..906D} analysis of corona dynamics of the open clusters, corona stars move along orbits close to periodic retrograde orbits, which may make the radial component of corona star velocity distribution close to zero. This means that the corona stars follow a spherical distribution and do not pass through the core of the cluster. We use the logarithmic Gaussian function to describe the RDP of the outer halo component, which also proves that the outer halo component has a spherical layer structure.

For ln$(r)=\mu$, the $g(r)=\rho$, represents the maximum number density of outer halo members.  We define the radius $r_o=r_{00} \cdot e^{\mu}$ as the mean size of the cluster's outer region, where $r_{00}$=1 degree. Around this place, for most of the clusters with outer halo components, the major component of cluster members has changed from core members to outer halo members. For comparison, within the cluster core radius $r_c$, the main component is the core members. In order to determine the boundary of star clusters, we define a radius $r_e=r_{00} \cdot e^{\mu+3\sigma}$ , which contains the majority of cluster members and can be considered as the boundary of the cluster. If $r=r_e$, then $g(r)=\rho \cdot e^{-9/2}$, the number density fraction is reduced to about 1\% of the maximum density of outer halo members, which also means $>$ 99.7\% outer halo members and $\sim$ 100\% core members are included within the $r_e$. 

\subsection{Fitting procedure}
After considering the distribution of outer halo members, we adopt the revised model with double components to perform a more reliable approximation of the cluster RDP and derive their characteristic radii. To perform a better RDP fitting, we divide the RDP into two parts ( the inner region and the outer region) and fit them separately. At first, we select a region whose radial number density is greater than half the maximum of the cluster RDP as the inner region and fit their RDP with the King model ( Equation~\ref{king} ). Since the core members are dominant in the inner region, the King model can always work well and presents a good approximation of the core member's RDP. In the outer region, because the fraction of outer halo members is increased along with the increase of radial distance, we subtract the distribution of core members according to the fitted King model and further use the logarithmic Gaussian function ( Equation~\ref{lognora}) to fit the RDP of the outer halo members. Finally, the defined characteristic radii  ( $r_c$, $r_t$, $r_o$, $r_e$ ) and the corresponding uncertainties ( $e\_r_c$, $e\_r_t$, $e\_r_o$, $e\_r_e$ ) were catalogued in the electronic table (see Section~\ref{sec:tab}).  

To provide reliable characteristic radii of open clusters, we plot the RDP of each cluster and check the fitting result by eyes. Although most open clusters (229 in 256) present the properties of two components in their RDP fitting, some of them (27 in 256) only present single component properties that can be approximately fitted by the King model. This means that the outer halo components of these clusters may have been stripped away by various internal and external factors in evolution, or they are not detected by us because they have a more extended and sparse outer halo structure. We then categorized them into two samples through visual inspection. In our catalog (see Table~\ref{tab1}), clusters that present the properties of two components in RDP are marked as the sample H (the flag labeled as 'H'), while clusters that can be fitted with the single King model are marked as the sample C (labeled as 'C'). 

\begin{figure*}
\centering
\includegraphics[scale=0.25]{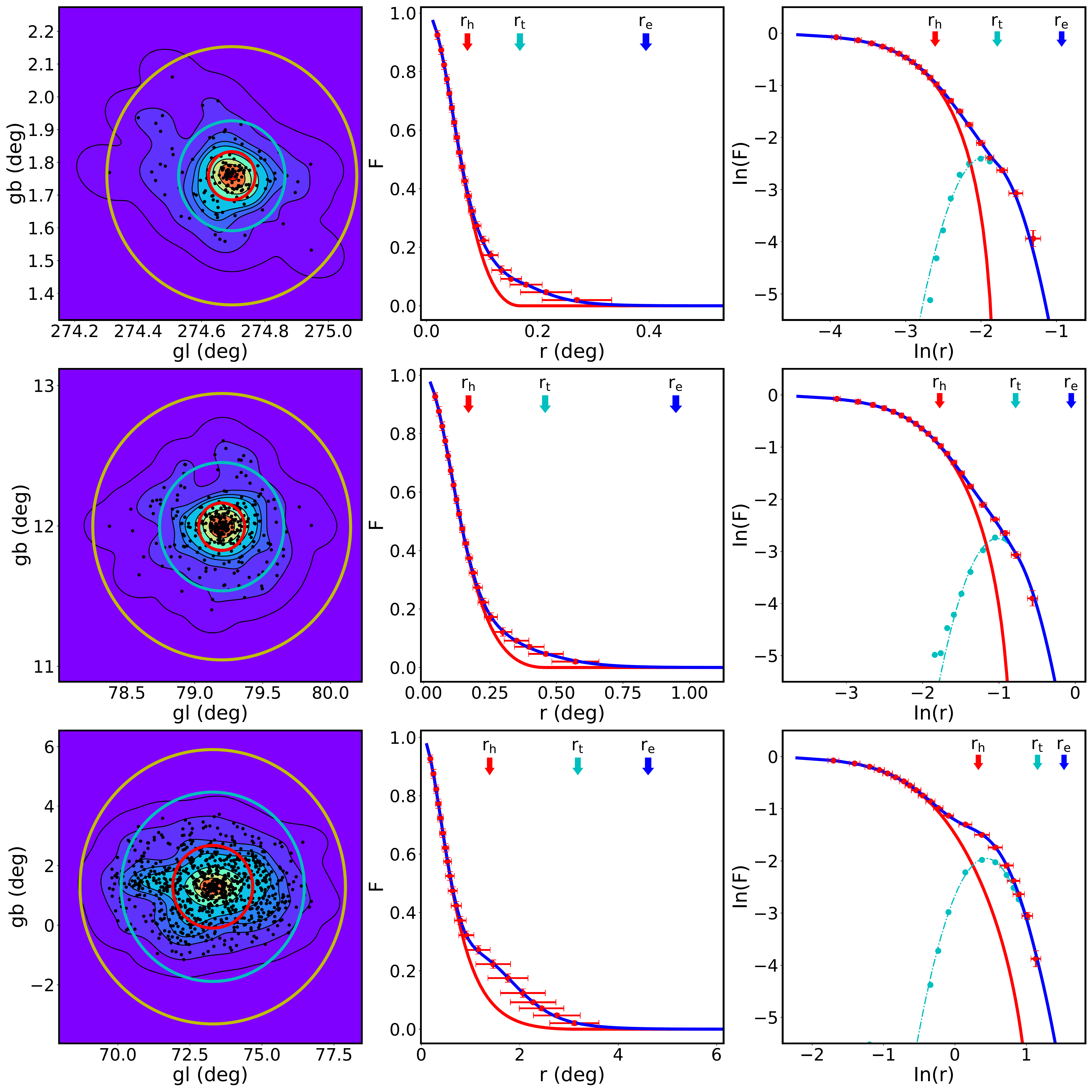}
\caption{Examples of three clusters: NGC\_2972 (upper panels), NGC\_6811 (middle panels), Gulliver\_17 (bottom panels). The left panels present the spatial distribution of member stars and their density contour. Three colors (red, cyan, yellow) are used to represent the spatial location of three radii ($r_h$, $r_t$, $r_e$). The middle panels present the RDP of cluster members (red dots). Red and blue lines represent fitting results of the King model and the double components model respectively. In the right panels, we plot the same RDP but with logarithmic coordinates. In addition to the core components (red lines), we also use cyan to represent the distribution and fitting results of the outer halo components of the cluster. Obviously, the double components model (blue lines) provides a good approximation for describing the RDP of star clusters.
}
\label{example}
\end{figure*}

Figure~\ref{example} shows the RDP of three typical clusters (NGC\_2972, NGC\_6811, and Gulliver\_17) as examples, while complete figures of 256 open clusters are provided in the electronic version of this paper. The left panels present the spatial distribution of all member stars and the density contour derived by the Gaussian kernel density estimation. Three circles with different colors ( red, cyan, yellow ) represent three kinds of radius ( $r_h$, $r_t$, $r_e$ ) respectively. It is obvious to note that: 1) within the $r_h$, because the core members are dominant, the spatial density distribution is dense and spherical symmetric; 2) between the $r_h$ and the $r_t$, because the number density of core members decreases rapidly ( the King profile ) and the number density of outer halo members increases gradually ( the logarithmic Gaussian function profile ), the spatial density distribution becomes loose and asymmetric; 3) outside the $r_t$ because most of the members are outer halo members, the spatial distribution presents extended, gradual, and asymmetric characteristics. In the middle panels, we plot the RDP of cluster members with red dots. Red and blue lines represent the King model $f(r)$ and our double components model $F(r)$ respectively.  For core members dominated within the $r_h$, their distribution is well consistent with the King profile. The large dispersion in the Y direction indicates that their density gradient is quite large, while the small dispersion in the X direction indicates that their spatial distribution is symmetric.  On the contrary, for outer halo members located outside the $r_h$, their distribution significantly deviates from the King model and the structure of core members: a small dispersion in the Y direction and a large dispersion in the X direction, which indicate the gentle density gradient and asymmetric spatial distribution of outer halo members. It is clear that the King model is no longer a good approximation after taking into account the distribution of outer halo members. Similarly, in the right panel, we also plot the RDP in logarithmic coordinate to better show the logarithmic Gaussian function of outer halo members with cyan dots and dotted lines. 

\begin{figure}
\centering
\includegraphics[scale=0.28]{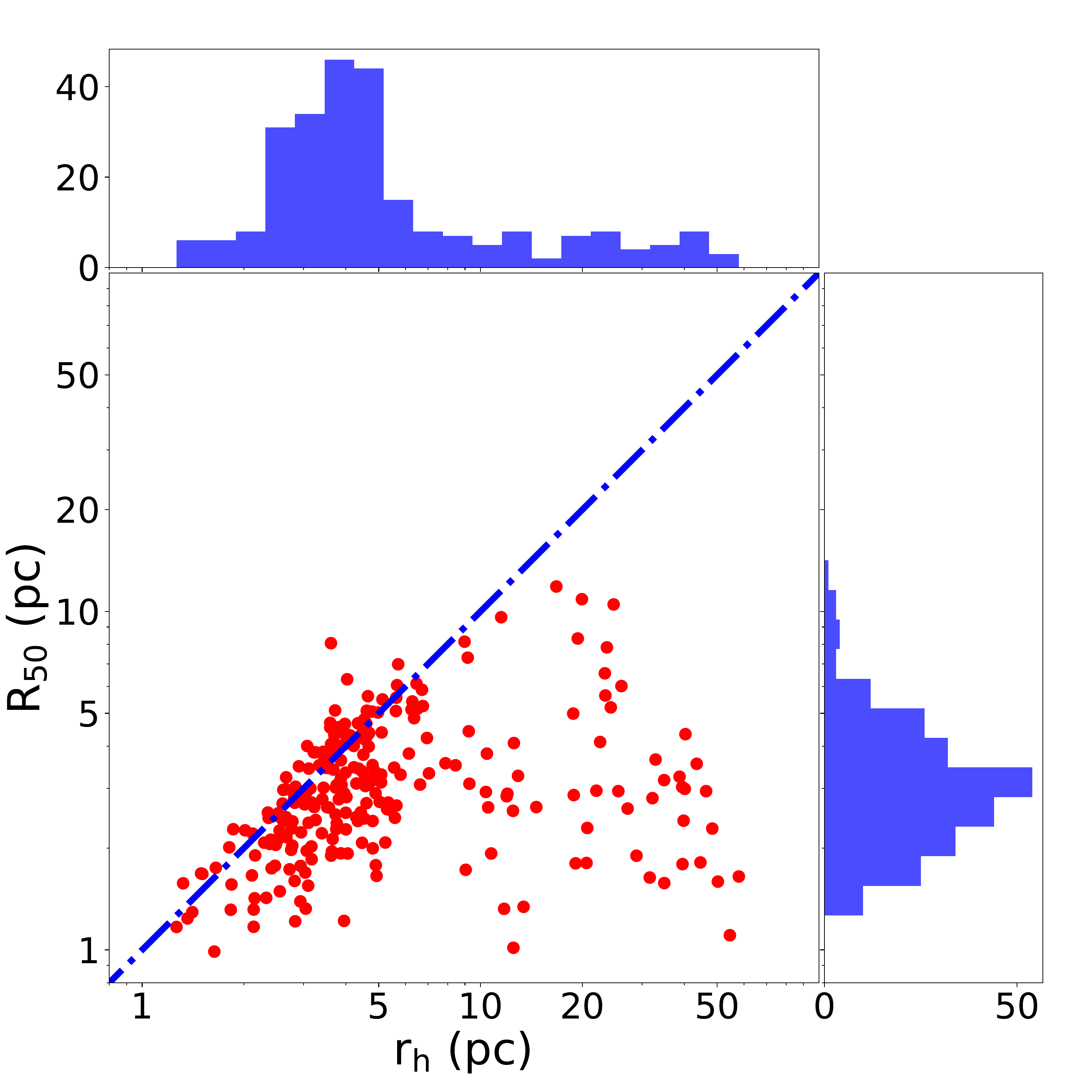}
\caption{Comparison of $R_{50}$ in CG20 and $r_h$ in this work. Since this work contains more member stars in the outer halo region, the $r_h$ presents a more extended distribution and is systematically larger than $R_{50}$ in CG20. }
\label{r50}
\end{figure}

\begin{figure*}
\centering
\includegraphics[scale=0.2]{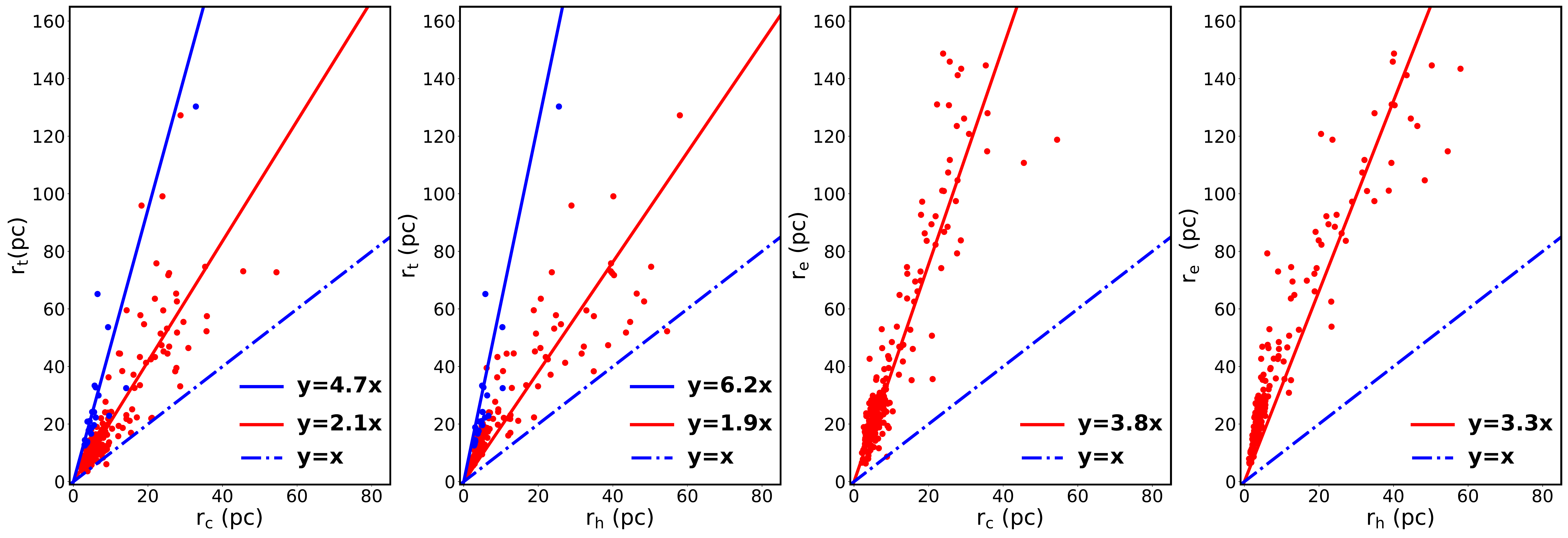}
\caption{Correlation relation with $r_t$ and $r_e$ as a function of $r_c$ and $r_h$. It is clear that all correlations can be approximated with a linear relation, which suggests a correlation between the internal and external components of the cluster. The blue dash-dotted lines are used to show the relation of y=x. In particular, we use blue dots and lines to show the radius correlation relation of clusters in sample C. The steeper slope in sample C indicates that these clusters may be affected by other dynamical processes (e.g., tidal stripping), compared with the clusters in sample H.}
\label{size}
\end{figure*}

\begin{figure*}
\centering
\includegraphics[scale=0.3]{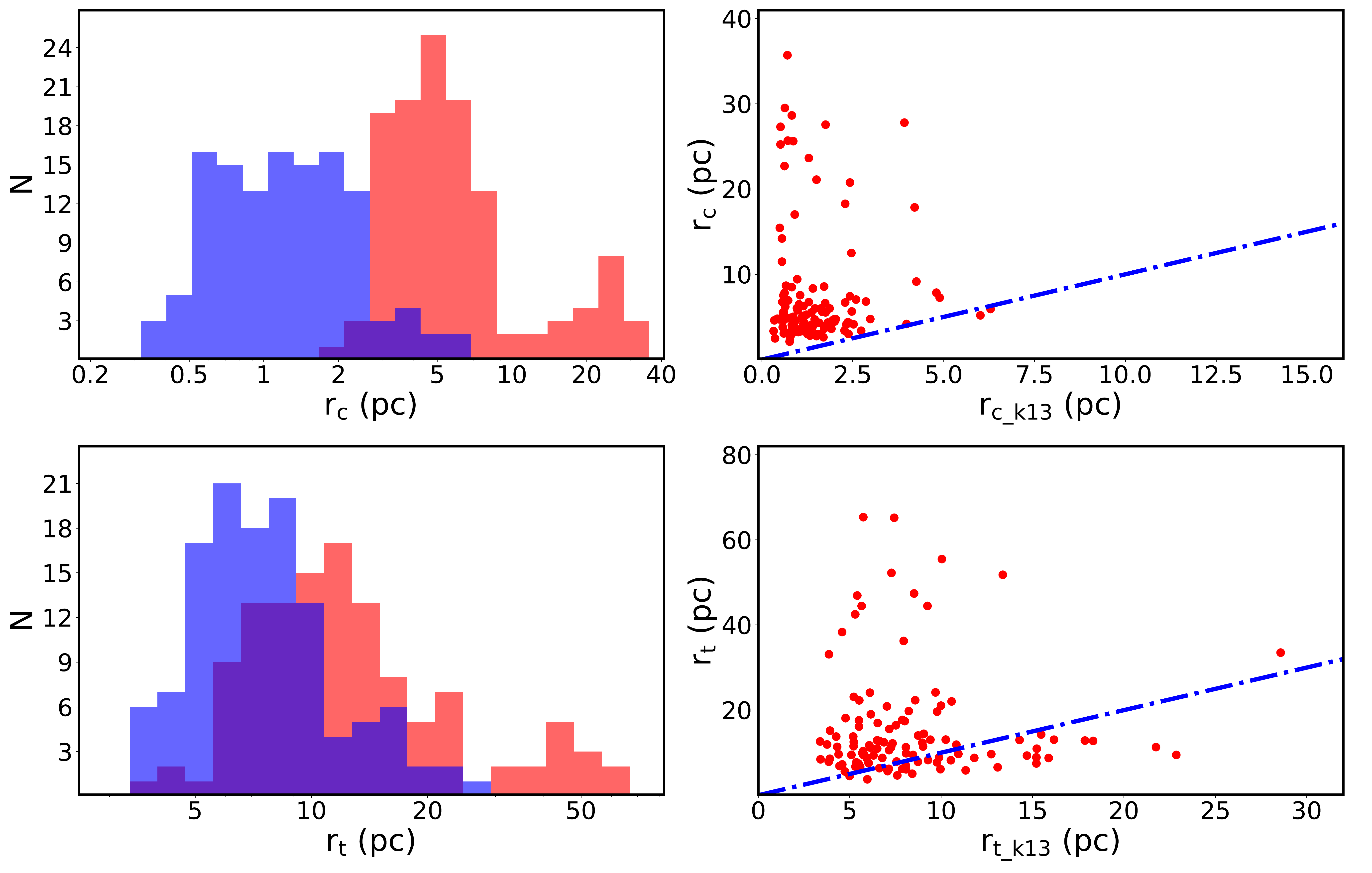}
\caption{Comparison of $r_c$ and $r_t$ between K13 and this work. \textit{Left  panels}: histogram distributions of $r_c$ and $r_t$ radius in K13 ( blue color) and this work (red color). \textit{Right panels}: comparison diagrams of $r_c$ and $r_t$ radius between K13 (x axis) and this work (y axis). The blue dash-dotted lines were used to show the relation of y=x. It is noted that, for most of the clusters, the $r_c$ in our catalog are systematically greater than in K13.}
\label{comparision}
\end{figure*}

\begin{figure}
\centering
\includegraphics[scale=0.24]{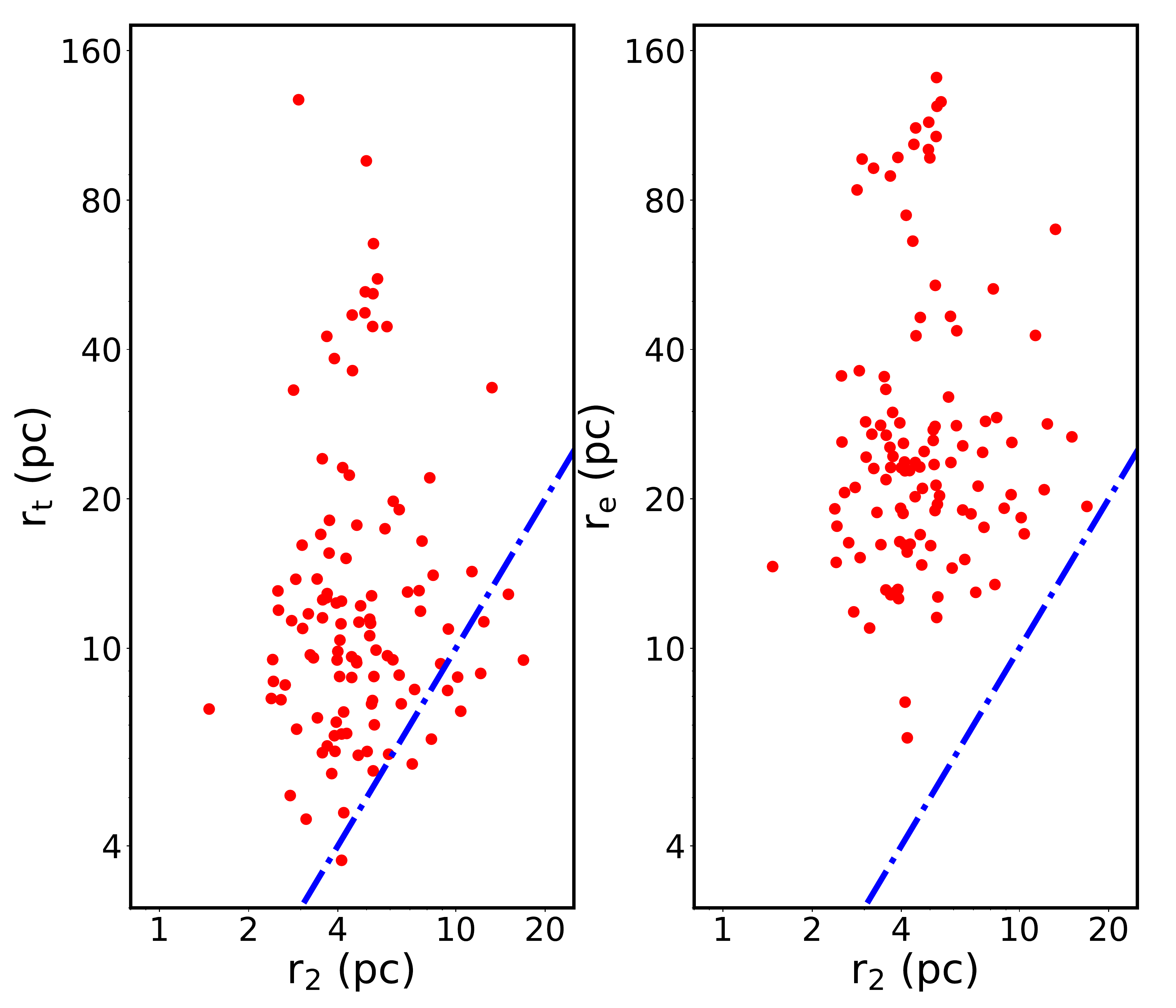}
\caption{Relationship between the cluster boundary $r_2$ in K13 and the boundary of two components in our catalog. We use blue dash-dotted lines to show the relation of y=x. Obviously, because of the existence of outer halo members, $r_e$ shows a larger size and can be regarded as the actual boundary of the open clusters. }
\label{r2}
\end{figure}

\begin{figure*}
\centering
\includegraphics[scale=0.35]{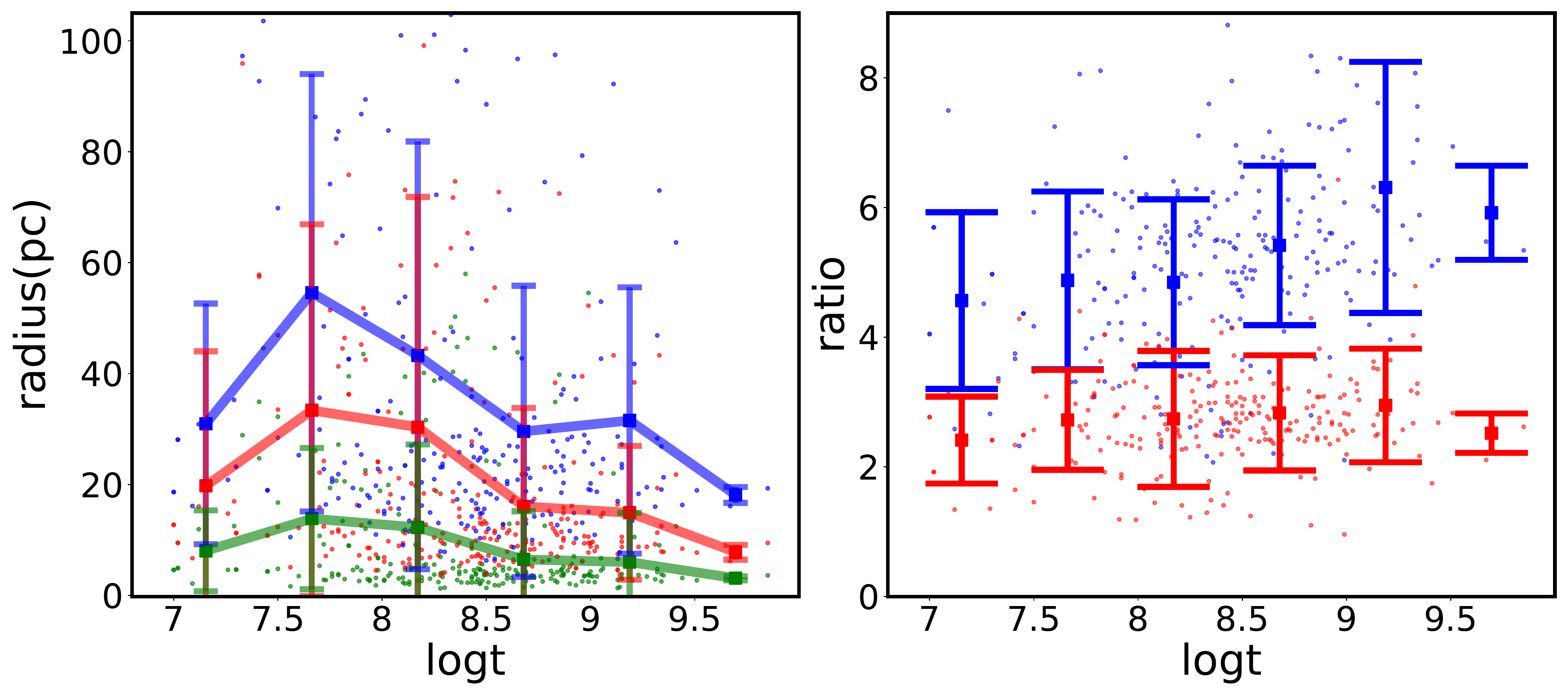}
\caption{Left panel: radii distribution of star clusters in different age bins. Green, red and blue colors are represent radii of $r_h$, $r_t$, $r_e$ respectively. Right panel: ratio of $r_t$/$r_h$ (red dots) and $r_e$/$r_h$ (blue dots) as function of star cluster ages. It is noted that the two panels only show the statistical relationship between the ages and radii of different star clusters, but can not be regarded as the evolution track of the single star cluster.}
\label{logt}
\end{figure*}

\section{Results and Discussion}
\label{sec:res}
\subsection{Open cluster radii}
In our study, the modified cluster RDP can be separated into two parts and adopted with four characteristic radii for description. The $r_c$ and $r_t$ represent the spatial distribution properties of core members which are mainly located on the inner part of a cluster, while $r_t$ represents the boundary of the core member components. The $r_o$ and $r_e$ describe an extended logarithmic Gaussian distribution which mainly represents the spatial properties of outer halo members. The $r_e$ radius not only indicates the boundary of the outer halo members but also represents the boundary of the whole cluster. 

In addition, we calculate a radius $r_h$ to better describe the cluster structure, which is defined as the location containing half of the identified members. The radius $r_h$ is independent of the RDP fitting and can be partially used to represent the physical size and concentration property of a cluster \citep{2020A&A...640A...1C, 2019ARA&A..57..227K}. Figure~\ref{r50} shows the comparison of $r_h$ between our work and CG20. In general, the $r_h$ from two catalogs presents a positive correlation relationship. However, since our catalog contains more member stars in the outer halo region, the $r_h$ in our catalog is systematically larger than in CG20. The mean values of half number radius in our catalogs and CG20 are 8.4 pc and 3.2 pc respectively. The underestimation of member stars in CG20  is also confirmed by \citet{2022A&A...659A..59T}.

To study the correlation and properties of the newly obtained characteristic radii, we convert the radii into their physical size with the distance provided by CG20, and plot Figure~\ref{size} for illustration. The radii distribution show that the size of $r_c$ and $r_h$ for the majority of clusters are less than 10 pc, with the median size of 5.8 pc and 4.1 pc respectively. This also means that both $r_c$ and $r_h$ represent a kind of similar radius size, the only difference is their derived method. We further study the correlation with $r_t$ and $r_e$ as a function of $r_c$ and $r_h$ and find that these correlation can be approximate with a linear relation y=a*x. The Pearson correlation coefficients of the four relations ($r_t$-$r_c$, $r_t$-$r_h$, $r_e$-$r_c$, $r_e$-$r_h$) are 0.61, 0.74, 0.87, 0.90 respectively. The linear relationship suggest an correlation between the internal and external components of the cluster, which also indicates that the spatial distribution of the two components may be related or driven by the same physical mechanism. Specifically, this means that the correlation can be used to expect the characteristic scale of star clusters. For example, if we only observe the cluster inner region and derived the core radius $r_c$ with the King model, we can further statistically predict the cluster boundary size $r_e$ as 3.8*$r_c$. 

In particular, we investigate the correlation relation of $r_t$-$r_c$ and $r_t$-$r_h$ of clusters belonging to sample C. In Figure~\ref{size}, we use blue dots and lines to represent this sample and find that their correlation has a steeper slope than that of sample H. For clusters in the two samples, we carefully compare the number of stars in the circle of radius r, which includes $r_c$, $r_h$, and $r_t$. We find that clusters in sample C have nothing special in their number density distribution. Therefore, we believe that the completeness of the two samples should be the same. For this reason, we prefer to suggest that the absence of halo components in sample C is due to the other physical factors (e.g., tidal stripping), rather than the incompleteness of observations.

In the right panel of Figure~\ref{size}, we note that the linear correlation of $r_e$ vs $r_h$ can not be simply fitted by a single slope. It seems that this correlation can be divided into two parts: small-size clusters with steeper slopes ($r_h$ $<$ 10 pc) and  large-size clusters with flatter slopes ($r_h$ $>$ 10 pc). Alternatively, we can also employ  second-order polynomial function to fit this correlation. However, taking into account different fitting functions may reflect different physical mechanisms in the formation and evolution of star clusters with different properties, we anticipate to study and discuss them in detail in the future.

\subsection{Comparison of open cluster radii}
In the study of the properties of star clusters, the cluster radius is of great significance to describe the structural characteristics of star clusters. As a widely cited star cluster catalog, K13 provide a catalog of 3006 Galactic star clusters and determine a homogeneous set of fundamental parameters. Based on the main morphological parts of a star cluster, three angular radii fitted by eye were provided by K13: $r_0$, $r_1$, $r_2$, which present the angular radius of the core, the angular radius of the central part, and the angular radius of the cluster respectively. According to the description in \citet{2012A&A...543A.156K}, for most of clusters, the $r_c$ is located between $r_0$ and $r_1$, and the $r_t$ is usually greater than $r_2$. In addition, K13 also provide the core radius $r_c$ and the tidal radius $r_t$ fitting with the King model.

In order to fully understand the characteristics of the radii we defined, we compare the newly obtained radii with the literature radii in K13. After cross-matching with our catalog and K13 catalog, we obtain 122 common clusters for comparison. Considering that K13 does not use KDE when performing the King model fitting process, different smoothing parameters and fitting methods will have a significant effect on the measurement results of $r_c$ and $r_t$. This may also be one of the reasons for the difference in the radius between the two catalogs in addition to the different number of member stars. 

Figure~\ref{comparision} shows radius comparison results of 122 common samples between K13 and this work. Although the $r_c$ of two catalogs were derived from the King model fitting, the clusters in our catalog have a larger size and contain more member stars, so it presents a larger core radius than in K13.  The mean size of $r_c$ in our catalog and K13 catalog is 7.7 pc and 1.5 pc respectively. However, for many clusters, the offset of $r_t$ is not very significant between K13 and this work (see the bottom panels in Figure~\ref{comparision}). This may reflect that base on the RDP of the core region, both works successfully reveal the boundary where the star density of cluster members becomes zero using the King model. The only difference worth noting is that, since the K13 cannot detect extended halo members beyond the $r_t$, the boundary of the cluster core component is regarded as the boundary of the entire cluster. With the help of Gaia astrometric data, we now know that the outer boundary of star clusters can actually extend much farther. Therefore, we use an additional extended outer halo component to better describe the structural characteristics of star clusters. In particular, we note that there are some clusters whose $r_c$ extends beyond 10 pc and $r_t$ extends beyond 40 pc. These star clusters may have experienced or are experiencing physical processes different from most star clusters. The nature of these clusters will be the focus of our future research. 

The actual visible radius of a cluster $r_2$ in K13 has long been cited as the outer boundary of star clusters in many literatures \citep[e.g.][]{2022ApJ...925..159Y,2021Ap&SS.366...92S,2018A&A...618A..93C}. However, in our work, most of the clusters can be divided into two components and their boundaries can be represent by two radii $r_t$ and $r_e$ respectively. Figure~\ref{r2} present the relation between $r_2$ in K13 and radius of $r_t$ and $r_e$ in our catalog. It is clear that the cluster boundary size estimated by either the core component ($r_t$) or outer halo component ($r_e$) are statistically greater than one expected before ($r_2$). The mean size of $r_2$, $r_t$ and $r_e$ of the 122 common samples are 5.2 pc, 21.6 pc and 31.6 pc respectively. Obviously, $r_e$ has a larger size and can be regarded as a new boundary of the cluster.  

\subsection{Cluster size vs. age}
We further discuss the radius scale distribution of clusters with different ages, trying to infer the evolution trend of cluster radius. Figure~\ref{logt} shows the radii distribution of clusters in different age bins. We use green, red, and blue to represent $r_h$, $r_t$, and $r_e$ respectively, which can be regarded as the size of the cluster central region, the size of the cluster core region, and the size of the cluster outer halo region. 

We find that the evolution trend of Figure~\ref{logt} can be roughly divided into three stages: 1) logt from 7 to 7.5, the cluster radius is gradually increasing, in which the increase of $r_e$ is the largest and $r_h$ is the smallest; 2) logt from 7.5 to 8.5, the cluster radius is decreasing, and $r_e$ decreases the most, especially in the period of logt from 7.5 to 8; 3) logt greater than 8.5, the radius of the cluster is roughly stable, and the mean values of $r_h$, $r_t$, and $r_e$ are 6.1 pc, 16.2 pc and 26.7 pc respectively. We ignore the result of logt $>$ 9.5 because there are too few cluster data in this age bin to reflect the statistics of old clusters. 

The internal evolution processes of star clusters mainly include mass loss due to stellar evolution and mass loss due to relaxation. For clusters with age $\sim$ 10-100 Myr, the stellar evolution has its greatest effects. Subsequently, the influence of mass loss caused by stellar evolution decreases gradually. At the same time, the relaxation-driven mass loss mechanism become the dominant factor in the internal evolution of star clusters, although the mass loss driven by this mechanism takes a much longer timescale\citep{2019ARA&A..57..227K}. The evolution trend we observed is generally consistent with the internal evolution of star clusters, including key time stages: 

- At cluster ages from 3 $\sim$ 40 Myr, due to the influence of supernova explosion, the cluster begins to lose lots of gas mass, which makes the gravitational potential well becomes shallow, and the cluster system becomes super-virial. The mass loss process makes the stars have an outward bulk motion, which leads to the increase of the cluster radii of all components. Because of the shallower potential well, the outer halo region loses more mass and lead to a more rapid expansion.

- For clusters with ages from 40 Myr to 1 Gyr, the mass loss of them will gradually be dominated by the envelope shedding of asymptotic giant branch (AGB) stars. Although this process is more gradual, it will lose 40 \% of the cluster's total mass during this stage. Furthermore, when more member stars escape to the distant region and become field stars, the size of the cluster is also gradually decreasing, especially the looser outer halo region;

- When the cluster age is greater than 1 Gyr, the mass loss due to stellar evolution gradually stops, while the total mass loss driven by the stellar evolution will not reach 50\% until ages on the order of a Hubble time. For old clusters ( $>$ 100 Myr), the relaxation-driven mass loss begins to dominate in the internal evolution of clusters and leads to slower mass loss.

It is worth noting that the evolutionary process discussed above mainly describes the cluster one can observe, while a large number of low mass star clusters may dissolute into field stars in their earlier evolution stage. Due to the continuous mass loss of star clusters in the evolution process, the older star clusters, the larger their initial mass and size. In other words, we can only regard Figure~\ref{logt} as the statistical relationship between the ages and radii of different star clusters, but not as the evolution track of a single star cluster. 

In the right panel of Figure~\ref{logt}, we present the ratio of $r_e$/$r_h$ (blue dots) and $r_t$/$r_h$ (red dots) as function of cluster ages (logt). Since $r_h$ represents the physical size of the cluster core component, and it does not change too much along with the cluster age,  this ratio can be used to measure the change of the relative size of the core component and outer halo component with cluster age. In general, the relative size of the core component keeps a constant, with mean and standard deviation being 2.6$\pm$0.2. Meanwhile, the relative size of the outer halo component increases slightly. Although the dispersion was large, the average ratio slowly increased from 4.4 to 6.3. Compared with young clusters, the ratio ($r_t$/$r_h$, $r_e$/$r_h$) of the old clusters shows that they still keep a dense core, but have a looser halo in the outer region. 

\begin{table*}
 \centering
\caption{Description of the open clusters catalog with structural parameters}
\label{tab1}
\begin{tabular}{llcl}
\hline
Column &  Format  & Unit   & Description \\
\hline
cluster & string & - & Cluster name in CG20 \\
ra & float & deg & Mean right ascension of members in CG20 \\
dec & float & deg & Mean declination of members in CG20 \\
pmra & float & mas yr$^{-1}$ & Mean proper motion along RA of members in CG20 \\
pmdec & float & mas yr$^{-1}$ & Mean proper motion along DE of members in CG20 \\
plx & float & mas & Mean parallax of members in CG20 \\
distpc & float & pc & Most likely distance of clusters in CG20 \\
logt & float & - & Estimated cluster age in CG20\\
bw & float & - & Bandwidth of kernel density estimation \\
$r_h$ & float & deg & Angular size of half number radius\\
$r_c$ & float & deg & Angular size of king's core radius derived by the core members \\
$e\_r_c$ & float & deg & Measurement uncertainty of King's core radius \\
$r_t$ & float & deg & Angular size of king's tidal radius derived by the core members \\
$e\_r_t$ & float & deg & Measurement uncertainty of King's tidal radius \\
$r_o$ & float & deg & Angular size of mean radius of outer halo members \\
$e\_r_o$ & float & deg & Measurement uncertainty of mean radius \\
$r_e$ & float & deg & Angular size of boundary radius of outer halo members \\
$e\_r_e$ & float & deg & Measurement uncertainty of boundary radius \\
$r_h\_{\rm pc}$ & float & pc & Physical size of half number radius\\
$r_c\_{\rm pc}$ & float & pc & Physical size of king's core radius derived by the core members \\
$r_t\_{\rm pc}$ & float & pc & Physical size of king's tidal radius derived by the core members \\
$r_o\_{\rm pc}$ & float & pc & Physical size of mean radius of outer halo members \\
$r_e\_{\rm pc}$ & float & pc & Physical size of boundary radius of outer halo members \\
flag & string & -& Label of cluster samples\\
  \hline
\end{tabular}
\end{table*}

\section{Description of the catalog}
\label{sec:tab}
In this paper, we provide a catalog to present the structure properties of 256 open clusters whose distance is between 1-2 kpc. The description of fundamental parameters of open clusters is listed in Table~\ref{tab1}, while the complete catalog of 256 open clusters is provided in the electronic version of this paper. Columns 1-8 list the basic parameters of clusters provided by CG20, including the cluster name, celestial coordinates, mean proper motions, parallax, distance, and age. Column 9 provides the bandwidth of kernel density estimation determined by Scott's rule, which could influence the estimate of the number density distribution of member stars. Columns 10-23 list the structural parameters derived in this paper, including the angular size, measurement uncertainty and physical size of $r_h$, $r_c$, $r_t$, $r_o$, $r_e$. In column 24, we provide a flag to mark the cluster whose RDP presents the outer halo component (labeled as 'H') or can be approximated as the single King model (labeled as 'C').

\section{Summary}
\label{sec:sum}
With the help of Gaia EDR3, we successfully re-determined member stars of 256 open clusters, especially for members located in the outer region of clusters. For most of these clusters ( 229 of 256 ), due to the existence of outer halo members, their RDP can no longer be simply approximated by the King model. Obviously, in the new insight of the open cluster, it not only has the high-density core component we observed before but also has an extended outer halo component hidden in the background of the field stars. Furthermore, there are many different characteristics in the spatial distribution of the two components.

To perform a better description of the cluster's RDP with an extended outer halo structure, we develop a revised model with double components for approximation: for the inner region which is dominated by the core members, the RDP is still mainly described by the King model; for the outer region, because the fraction of outer halo members increases gradually from inside to outside, we adopt the logarithmic Gaussian function to describe the deviation of the cluster's RDP from the King model. In other words, the spatial structure of outer halo members can be well approximated with the logarithmic Gaussian function. 
After fitting the RDP of clusters with our double component model, we define four characteristic radii for describing the cluster structure: in the inner part, the $r_c$ and $r_t$  which derived from the King model represent the spatial distribution properties of core members; in the outer part, the $r_o$ and $r_e$ which derived from the logarithmic Gaussian function represent the spatial properties of outer halo members. Although these four radii describe the structural characteristics of different components of the cluster, there is a good linear relationship between them, which shows that the structure of two components is still a whole and follows the same physical law. We note that although $r_h$ does not depend on the structural model assumption of the cluster, but its size is similar to $r_c$, which means that it is also appropriate to use $r_h$ to describe the structural characteristics of the cluster. According to our correlation results (see Figure~\ref{size} ), $r_h$ can also be used to simply track the approximate region and distribution of core components and outer halo components.

{\bf Acknowledgments}
We are very grateful to the anonymous referee for helpful information and suggestions. This work is supported by National Key R\&D Program of China No. 2019YFA0405501. Li Chen acknowledges the support from the National Natural Science Foundation of China (NSFC) through the grants  12090040 and 12090042. Jing Zhong would like to acknowledge the NSFC under grants 12073060, and the Youth Innovation Promotion Association CAS. We acknowledge the science research grants from the China Manned Space Project with NO. CMS-CSST-2021-A08.
This work has made use of data from the European Space Agency (ESA) mission GAIA (\url{https://www.cosmos.esa.int/gaia}), processed by the GAIA Data Processing and Analysis Consortium (DPAC,\url{https://www.cosmos.esa.int/web/gaia/dpac/consortium}). Funding for the DPAC has been provided by national institutions, in particular the institutions participating in the GAIA Multilateral Agreement.

\end{CJK*}
\end{document}